\documentclass[conference]{IEEEtran}
\IEEEoverridecommandlockouts
\usepackage{cite}
\usepackage{amsmath,amssymb,amsfonts}
\usepackage{algorithmic}
\usepackage{graphicx}
\usepackage{textcomp}
\usepackage{xcolor}
\usepackage{xspace}
\usepackage{booktabs}
\usepackage{caption}
\def\BibTeX{{\rm B\kern-.05em{\sc i\kern-.025em b}\kern-.08em
    T\kern-.1667em\lower.7ex\hbox{E}\kern-.125emX}}

\newcommand{\sysname}{{\it EarDA}\xspace}

\begin{document}

\title{\sysname: Towards Accurate and Data-Efficient Earable Activity Sensing}

\author{
    \IEEEauthorblockN{
        Shengzhe Lyu,
        Yongliang Chen,
        Di Duan,
        Renqi Jia,
        Weitao Xu\IEEEauthorrefmark{1}
    }
    \IEEEauthorblockA{
        \textit{Department of Computer Science, City University of Hong Kong} \\
        Hong Kong SAR, China
    }
    \thanks{* Corresponding author.}
}


\maketitle
\begin{abstract}
  In the realm of smart sensing with the Internet of Things, earable devices are empowered with the capability of multi-modality sensing and intelligence of context-aware computing, leading to its wide usage in Human Activity Recognition (HAR). Nonetheless, unlike the movements captured by Inertial Measurement Unit (IMU) sensors placed on the upper or lower body, those motion signals obtained from earable devices show significant changes in amplitudes and patterns, especially in the presence of dynamic and unpredictable head movements, posing a significant challenge for activity classification. In this work, we present \sysname, an adversarial-based domain adaptation system to extract the domain-independent features across different sensor locations. Moreover, while most deep learning methods commonly rely on training with substantial amounts of labeled data to offer good accuracy, the proposed scheme can release the potential usage of publicly available smartphone-based IMU datasets. Furthermore, we explore the feasibility of applying a filter-based data processing method to mitigate the impact of head movement. \sysname, the proposed system, enables more data-efficient and accurate activity sensing. It achieves an accuracy of 88.8\% under HAR task, demonstrating a significant 43\% improvement over methods without domain adaptation. This clearly showcases its effectiveness in mitigating domain gaps.
\end{abstract}


\section{Introduction}
Earable devices, which significantly enhance people's everyday listening experience, have experienced a noticeable surge in recent years.
With technological advancements, many newly launched devices have been integrated with various sensors (e.g., AirPods Pro 2, Bose QuietControl 30 and Sony WF-1000XM5), giving rise to an emerging research area known as earable sensing.
Compared to other sensing modalities, earable sensing holds a significant advantage as it seamlessly integrates into users' daily lives.
Consequently, it has attracted researchers' attention and has been shown to be effective in various tasks, including health care~\cite{bedri2017earbit, roddiger2019towards}, speech enhancement~\cite{duan2024earse, he2023towards} and activity recognition~\cite{hossain2019human, kawsar2018earables, burgos2020ear}.

Existing works in this realm are either developed based on dedicated or commercial off-the-shelf (COTS) devices.
The first type requires specially designed devices, such as eSense~\cite{kawsar2018esense} and OpenEarable~\cite{roddiger2022openearable}.
Despite their effectiveness, the requisite highly customized hardware significantly restricts their practicability.
On the other hand, with the emergence of sensor-rich earable products like Apple Airpods, an increasing number of researchers have proposed plausible solutions based on these COTS devices~\cite{duan2024earse, li2023echoattack}.
However, the obtained sensory data is highly heterogeneous across different manufacturers, and there are no unified datasets applicable to all devices, which suggests a dilemma where developing new sensing applications often necessitates collecting the whole dataset from scratch.
As such, \textit{how to develop an accurate earable sensing methodology in a data-efficient manner based on COTS devices} remains an unsolved ad-hoc question.

To fill this research gap, in this paper, we propose \sysname, an earable human activity recognition (HAR) system based on domain adaption. 
To the best of our knowledge, we are the first to propose such a system that can utilize knowledge learned from other sensing modalities (e.g., smartphones) with abundant training data to enable effective earable-based HAR with limited data.
However, directly applying well-trained frameworks from other modalities to earable data is impracticable due to the significant domain gap that exists between these two modalities (e.g., smartphones are always carried in pockets while earable devices are attached to the user's head).
Thereafter, we specially design a domain adaptation training framework to leverage HAR features learned from public datasets with substantial sensory data to boost the performance on new earable devices.
It effectively reduces the required human labor intensity on collecting abundant earable training data due to the absence of unified datasets.
In addition, we encounter another challenge that head movements introduce distinct signal variations, which overwhelm the informative signal for activity recognition.
To mitigate this, we carefully investigate the obtained signals in different scenarios and devise a filter-based solution that effectively mitigates the resulting effects.

We have built a prototype of \sysname atop Apple AirPods Pro 2, and it is evaluated on four common daily activities (walking, upstairs, standing, and jogging) based on IMU measurements.
The training is conducted with a small volume of earable data (i.e., 5 minutes) with the assistance of four public smartphone-based IMU datasets.
It recognizes activities at an accuracy of 89\%, with a significant increment of 43\%, compared with cases only training with the public dataset and adopting the model on earables without domain adaption, demonstrating the proposed system's effectiveness. Incorporating a filter-based data processing method to mitigate head movement enhances the overall accuracy by 4\%, highlighting the system's ability to cooperate with the head interference. In short, we make the following contributions:
\begin{itemize}
    \item We take the first step toward understanding the domain gap existing in the public wearable datasets, especially smartphone-based datasets and IMU data from COTS earables.
    \item We propose \sysname with novel techniques that mitigate the domain gap between COTS earable devices and public datasets and influences introduced by irregular head movements.
    \item We conduct extensive evaluations and showcase the superiority of the proposed system. The overall accuracy achieves 88.8\%, and the ablation study demonstrates the effectiveness of the proposed components.
\end{itemize}

\section{Feasibility Study}
\subsection{Impact of Location Diversity}
\begin{figure*}[t]
  \centering
  \includegraphics[width=\textwidth]{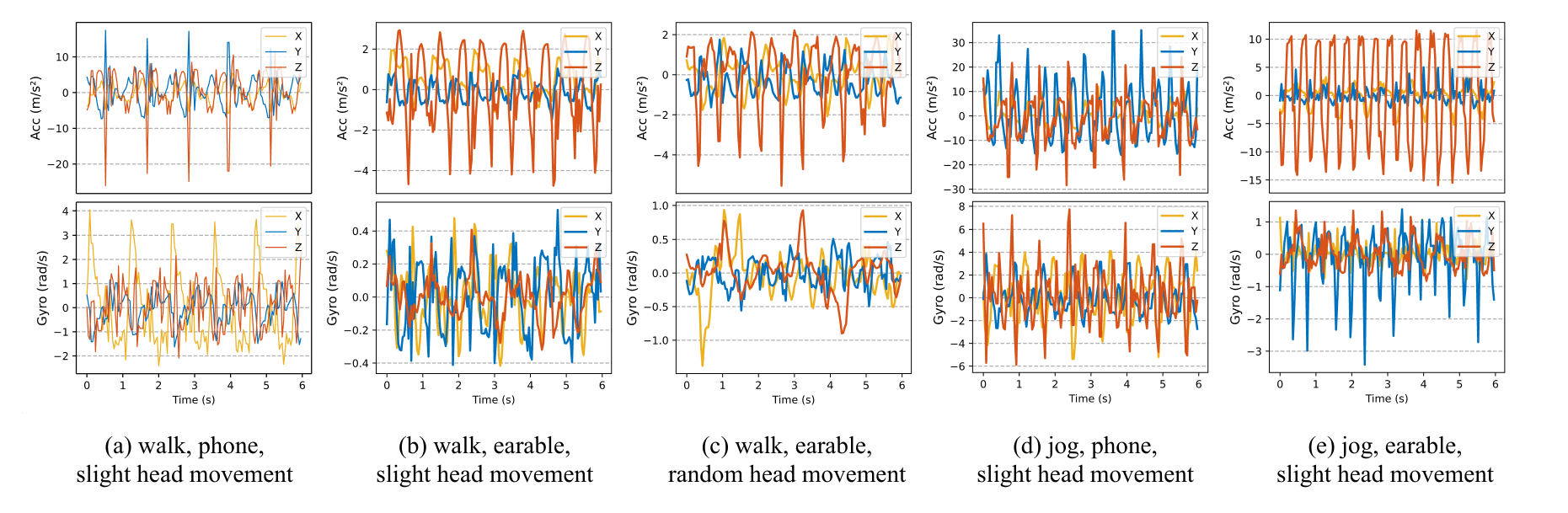}
  \vspace{-0.4in}
  \caption{IMU measurements by smartphone (left pocket) and earable (head) for different activities and head movements.}
  \label{fig:imu_sequence}
\end{figure*}

The disparity among sensory data collected by different modalities mainly stems from the different sensing locations. To better illustrate such discrepancies, we collect IMU data from a smartphone placed in the pocket of the trousers and a naturally-worn earable device when performing different actions. We visualize their sensory data in Figure~\ref{fig:imu_sequence}.
It can be observed from Figure~\ref{fig:imu_sequence}(a) and Figure~\ref{fig:imu_sequence}(b) that there are notable differences in both of the acceleration and gyroscope measurements.
The smartphone's sensor records the movements and vibrations of the lower body, reflecting the motion patterns and gestures associated with leg and hip movements. On the other hand, the earable device captures a combination of head and body movements, providing a different perspective on overall body motion. For example, the maximum range of acceleration during jogging is observed to be twice that of earables, as shown in Figure~\ref{fig:imu_sequence}(d, e). When measuring acceleration signals in the lower body, the amplitude of the signals tends to be larger.
Nevertheless, apart from all the challenges, the presence of similar patterns, such as periodic peaks in both domains, underscores the potential to transfer the knowledge across domains.

\subsection{Impact of Head Movement}
Head movement is unavoidable when performing daily activities. However, the introduced signal fluctuations pose extra challenges in sensing tasks. As shown in Figure~\ref{fig:imu_sequence}(c), even when the lower body and upper body are experiencing periodical movements during walking, unpredictable interference appears in IMU signals collected by the ear-worn device due to the head movement. In addition, the impact of head movements is still obvious while the users tried their best to control their heads during the data collection, as shown in Figure~\ref{fig:imu_sequence}(b). This can significantly affect the HAR tasks because the noise and variations distort the IMU data patterns,
making it more difficult to accurately classify the user's activity.
Fortunately, due to the filtering function of the upper body, the frequency components of earable IMU data always fall in a certain range, suggesting the usage of filter-based data processing methods might help to mitigate the interference from head movements.

\section{System Design}
\begin{figure*}[t]
  \centering
  \includegraphics[width=0.65\textwidth]{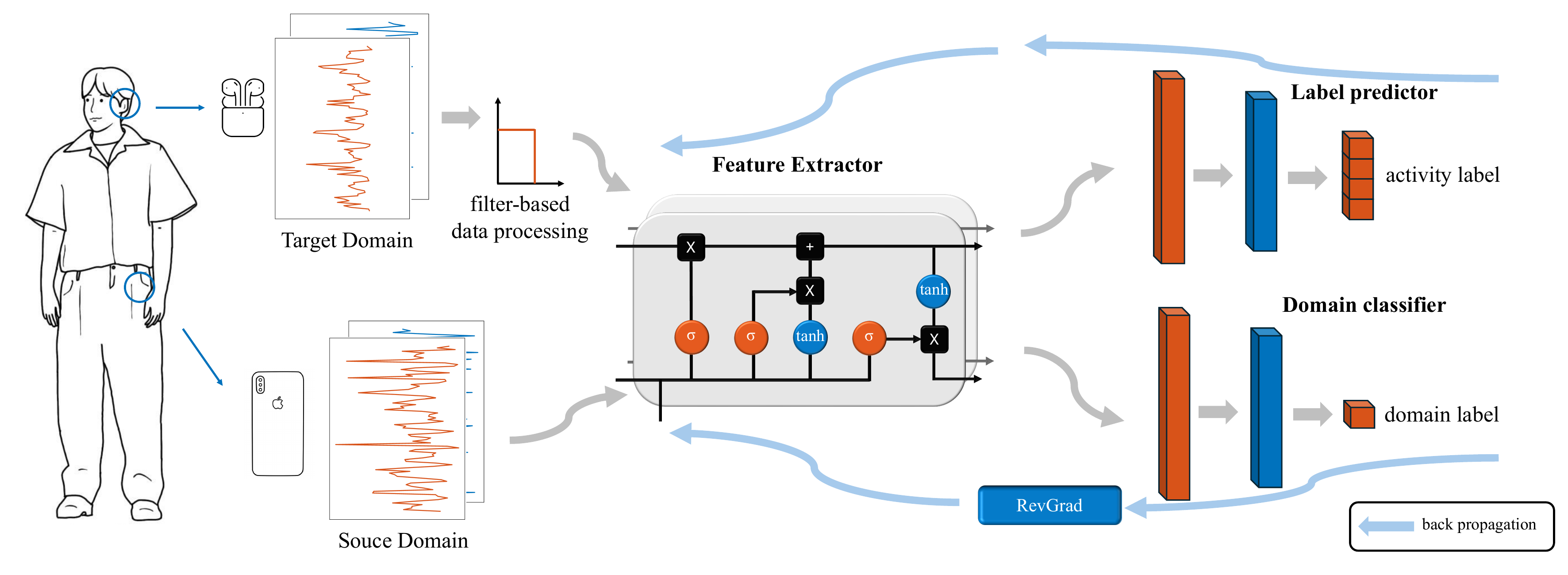}
  \vspace{-0.2in}
  \caption{System overview.}
  \vspace{-0.15in}
  \label{fig:system_overview}
\end{figure*}
\subsection{System Overview}
The overview of \sysname is shown in Figure~\ref{fig:system_overview}.
We design an effective adversarial-based domain adaptation module that can extract domain-invariant features across the source domain with substantial data (i.e., public datasets) and the target domain with limited data (i.e., earable data).
Moreover, we specially design a filter-based data pre-processing pipeline to mitigate the influence caused by head movements.

\subsection{Domain-invariant Feature Extraction}
\paragraph{Feature Extractor}
A bi-directional LSTM model is designed to construct a generalized feature extractor due to its architecture, which is adapted to capturing the complex temporal dynamics and dependencies inherent in sequential data. To boost the model's robustness against device orientation and data heterogeneity, the acceleration and gyroscope magnitudes are calculated as inputs. It serves as a pre-processing step to abstract away the device's orientation, providing a more orientation-invariant representation of the motion. This approach enhances the model's generalization ability across different devices and domains. It can be seen in Figure~\ref{fig:imu_sequence} that the acceleration readings are always steadier than gyroscope readings during the same activity. The values derived from the accelerometer and gyroscope always fall on different ranges as the gyroscope is more sensitive to movement. To mitigate the impact, accelerations are normalized through the division by gravity, after which the range differences between acceleration readings and gyroscope readings can be narrowed.

\paragraph{Domain Adaptation}
As mentioned previously, the normal LSTM classifier trained on the source domain cannot be directly applied to the target domain. As a result, the model should retain the capability to extract domain-invariant features regardless of which domain the input data falls on. In this work, we present a domain adaptation module that takes substantial data from the source domain and limited data from the target domain as input. Then, two distinct pathways are derived from the feature extractor, one label predictor, and one domain classifier. The label predictor is tasked with the objective of predicting the correct activity label for the input IMU sequence. In contrast, the domain classifier, aided by the gradient reversal layer, attempts to determine the domain of the input data.

Inspired by~\cite{ganin2015unsupervised}, we integrate a Gradient Reversal Layer (GRL) to assist in learning domain-invariant features of the two domains. Specifically, during the forward pass, the GRL acts as an identity function, allowing data to pass through unchanged. However, during the backward pass, it reverses the gradient sign before passing it back to the preceding layers. This effectively encourages the LSTM layers to learn representations that are indistinguishable from the domain classifier. Consequently, this process leads to the minimization of domain discrepancy, resulting in the capability of domain-invariant feature extraction.
More formally, the loss function for the network utilizing GRL can be written as:
\begin{align*}
    E(\theta_f, \theta_y, \theta_d) 
    &= \sum_{\substack{i=1..N}} L_y \left( G_y(G_f(x_i;\theta_f);\theta_y), y_i \right) \\
    &- \lambda \sum_{i=1..N} L_d \left( G_d(G_f(x_i;\theta_f);\theta_d), y_i \right) \\
    &= \sum_{\substack{i=1..N}} L_y^i(\theta_f, \theta_y) - \lambda \sum_{i=1..N} L_d^i(\theta_f, \theta_d)
\end{align*}
Here, $G_f, G_y, G_d$ are the functions representing the feature extractor, label predictor, and domain classifier over the parameters of $\theta_f, \theta_y, \theta_d$, respectively. $\lambda$ is a trade-off parameter that controls the importance of the domain adaptation loss relative to the label prediction loss, which is selected to be 0.3 in our work. $L_y$ is the label predictor loss and $L_d$ is the domain classifier loss. This loss function aims to minimize the label prediction error while also encouraging the feature extractor to learn domain-invariant features. The presence of the GRL is implied in the negative sign before the domain classifier loss term, which reflects the gradient reversal during the backpropagation process.

\subsection{Mitigation of Head-movement}
\begin{figure*}[t]
  \centering
  \begin{minipage}[t]{0.48\linewidth}
    \centering
    \includegraphics[width=0.9\linewidth]{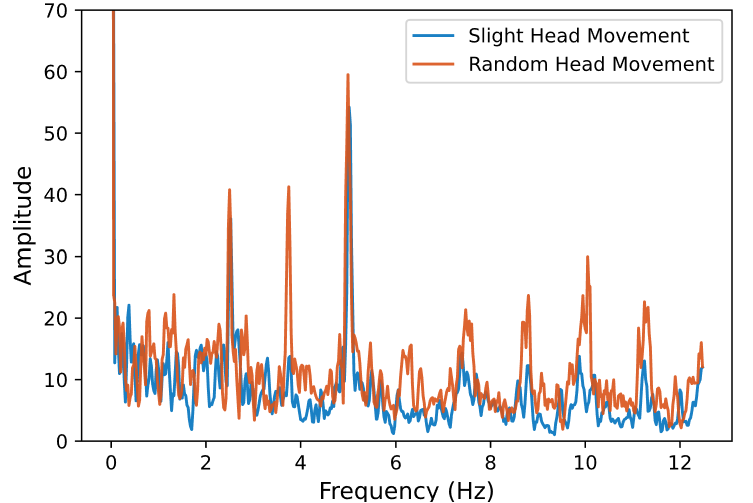}
    \caption{Frequency spectrum of gyroscope measurements collected by earables during jogging.}
    \label{fig:fft}
  \end{minipage}\hfill
  \begin{minipage}[t]{0.5\linewidth}
    \centering
    \includegraphics[width=\linewidth]{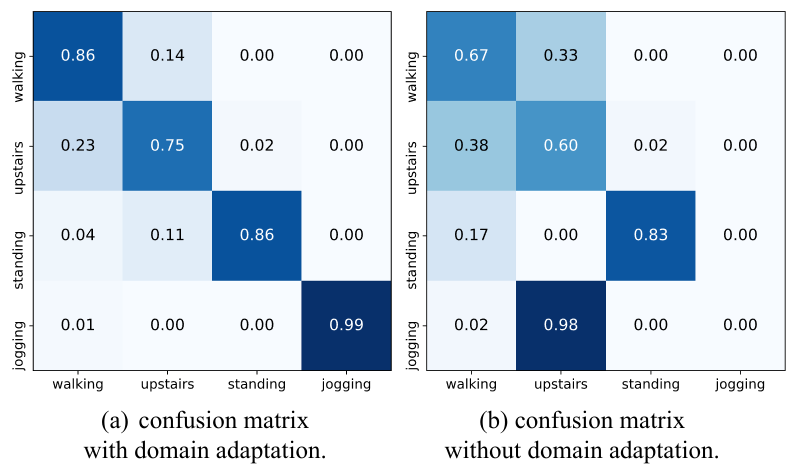}
    \caption{Confusion matrix on the test set with and without applying domain adaptation.}
    \label{fig:confusion_matrix}
  \end{minipage}
\end{figure*}

The head movement interference has been a challenge in performing HAR tasks with earables because the signals recorded by ear-worn devices are usually unpredictable with the interference added.
One approach to mitigate the impact of head movements on HAR tasks is to employ signal processing like applying filters.
It must be admitted that eliminating the motion caused by head movements from raw IMU data is not easy, as the frequencies of body or head movements spread over a big range and overlap each other in a specific range. As studied in~\cite{antonsson1985frequency}, human activity frequencies are always between 0 and 20 Hz and 98\% of the frequency spectrum is below 10 Hz. 
Nonetheless, due to the filtering function of the body structure, the signals recorded on the head are usually smaller, always below 5.8 Hz, presented in~\cite{rinaudo2019human}. This leads to the potential of mitigating head movement by low-pass filtering.
We observed similar frequency component distribution by comparing the frequency spectrum under different kinds of head movements.
As illustrated in Figure~\ref{fig:fft}, the dominant frequency components of gyroscope data are concentrated in low-frequency bins (e.g., below 5 Hz). However, when the same activity is accompanied by more rigorous head movements, noticeable lobes appear at higher frequency ranges.
Considering previous literature on regular human activities~\cite{antonsson1985frequency,rinaudo2019human} and the observed sensory data, we empirically devise a low-pass filter with a cut-off frequency of 5 Hz.
It can effectively filter out unwanted disturbances occupying high-frequency ranges, including those introduced by erratic head movements, while simultaneously retaining informative signals related to body movements.

\section{Evaluation}
\subsection{Implementation}
To better understand the impact of sensor locations and head movements on the performance of HAR tasks, we collected a dataset with Apple Airpods Pro 2. The Airpods are chosen because: 1) the APIs are provided to access accurate and real-time 6-axis motion data captured from the accelerometer and gyroscope embedded in the earbuds, 2) IMU signals from each earbud in a pair are fused together automatically in API to represent the motion signals of the head, and 3) it is widely utilized in HAR tasks on earable devices as it is one of the most popular COTS earable products. 
The sampling rate of Airpods was set to the highest value allowed by API, which is 25 Hz.

A two-layer bidirectional LSTM model with a hidden size of 16 is implemented as the feature extractor. The model takes two-dimensional inputs with a sequence length of 100 from each domain. Fully connected layers with ReLU activation are employed for both the label predictor and the domain classifier. The gradient reversal layer is only implemented as part of the domain classifier. The model was trained with a batch size of 32 for 200 epochs on the computer with an NVIDIA GeForce RTX 2080Ti GPU.

\subsection{Experimental Setup}
\subsubsection{Earable Data Collection}
\begin{figure}[t]
  \centering
  \includegraphics[width=1\linewidth]{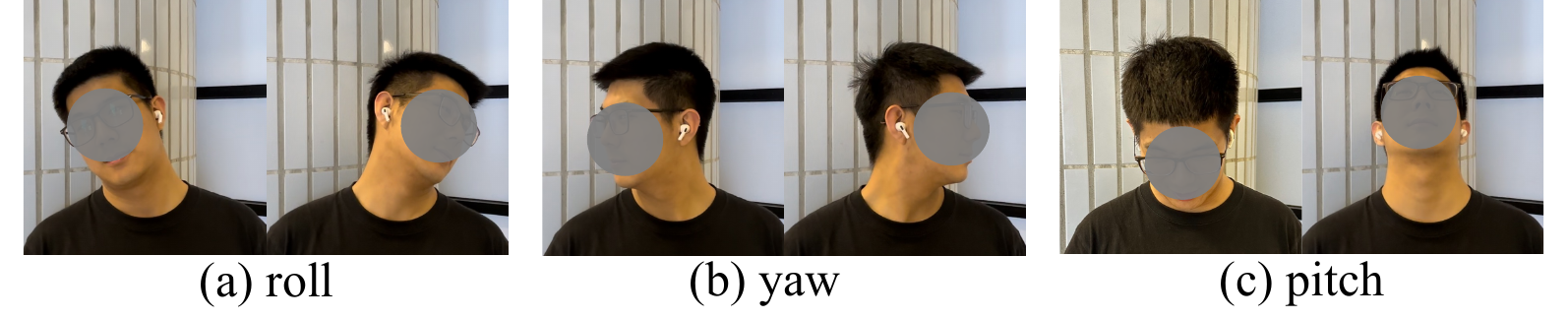}
  \vspace{-0.2in}
  \caption{Three evaluated head movements in the experiment.}
  \vspace{-0.15in}
  \label{fig:earables}
\end{figure}

\begin{table*}
    \centering
    \caption{Performance of the model with different head movements after mitigation of head movements.}
    \label{tab:model_performance_gr_filter}
    \renewcommand{\arraystretch}{0.9}
    \begin{tabular}{ccccccccccccc}
    \toprule
    \textbf{Head movement} 
    & \multicolumn{2}{c}{\textbf{Slight}}
    & \multicolumn{2}{c}{\textbf{Random}} 
    & \multicolumn{2}{c}{\textbf{Roll}} 
    & \multicolumn{2}{c}{\textbf{Yaw}} 
    & \multicolumn{2}{c}{\textbf{Pitch}} 
    & \multicolumn{2}{c}{\textbf{Average}} \\ \midrule
    \textbf{Metric} & \centering\textbf{Acc} & \textbf{F1} & \textbf{Acc} & \textbf{F1} & \textbf{Acc} & \textbf{F1} & \textbf{Acc} & \textbf{F1} & \textbf{Acc} & \textbf{F1} & \textbf{Acc} & \textbf{F1} \\ \midrule 
    \textbf{Walking} & 0.90 & 0.82 & 0.76 & 0.83 & 1.00 & 0.63 & 0.92 & 0.92 
    & 0.88 & 0.92 & 0.82 & 0.82 \\
    \textbf{Upstairs} & 0.77 & 0.81 & 0.89 & 0.79 & 0.62 & 0.76 & 0.80 & 0.89
    & 0.86 & 0.92 & 0.78 & 0.81 \\
    \textbf{Standing} & 0.92 & 0.96 & 1.00 & 1.00 & 1.00 & 0.96 & 1.00 & 0.86 
    & 1.00 & 0.86 & 0.97 & 0.94 \\
    \textbf{Jogging} & 1.00 & 0.99 & 1.00 & 1.00 & 1.00 & 1.00 & 1.00 & 1.00
    & 1.00 & 1.00 & 1.00 & 0.99 \\
    \midrule
    \textbf{Overall} & \textbf{0.90} & \textbf{0.89} &\textbf{0.91} & \textbf{0.91} & \textbf{0.90} &\textbf{0.84} &\textbf{0.93}
    & \textbf{0.92} & \textbf{0.94} & \textbf{0.93} &\textbf{0.89} & \textbf{0.89} \\
    \bottomrule
\end{tabular}
\end{table*}
\begin{table*}
    \centering
    \caption{Performance of the model with different head movements.}
    \label{tab:model_performance_gr}
    \renewcommand{\arraystretch}{0.9}
    \begin{tabular}{cccccccccccccc}
    \toprule
    \textbf{Head movement} 
    & \multicolumn{2}{c}{\textbf{Slight}} 
    & \multicolumn{2}{c}{\textbf{Random}}
    & \multicolumn{2}{c}{\textbf{Roll}}
    & \multicolumn{2}{c}{\textbf{Yaw}}
    & \multicolumn{2}{c}{\textbf{Pitch}} 
    & \multicolumn{2}{c}{\textbf{Average}} \\ \midrule
    \textbf{Metric} & \textbf{Acc} & \textbf{F1} & \textbf{Acc} & \textbf{F1} & \textbf{Acc} & \textbf{F1} & \textbf{Acc} & \textbf{F1} & \textbf{Acc} & \textbf{F1} & \textbf{Acc} & \textbf{F1} \\ \midrule 
    \textbf{Walking} & 0.78 & 0.86 & 0.76 & 0.80 & 0.86 & 0.39 & 0.87 & 0.67 
    & 1.00 & 0.59 & 0.80 & 0.73 \\
    \textbf{Upstairs} & 0.95 & 0.82 & 0.83 & 0.78 & 0.52 & 0.69 & 0.52 & 0.68
    & 0.50 & 0.67 & 0.64 & 0.72 \\
    \textbf{Standing} & 0.98 & 0.99 & 1.00 & 1.00 & 1.00 & 0.88 & 1.00 & 0.70 
    & 1.00 & 0.74 & 0.99 & 0.91 \\
    \textbf{Jogging} & 1.00 & 0.99 & 1.00 & 1.00 & 1.00 & 1.00 & 1.00 & 1.00
    & 1.00 & 1.00 & 1.00 & 0.99 \\
    \midrule
    \textbf{Overall} & \textbf{0.93} & \textbf{0.92} & \textbf{0.90} &\textbf{ 0.90} &\textbf{ 0.84} & \textbf{0.74} & \textbf{0.85}
    & \textbf{0.76} & \textbf{0.88} & \textbf{0.75} & \textbf{0.86} & \textbf{0.84} \\
    \bottomrule
\end{tabular}
\vspace{-0.1in}
\end{table*}

During the data collection with Airpods, the participant was asked to perform four different activities \textit{(walking, walking upstairs, standing, jogging)} while wearing the Airpods as usual. The total time for each activity is 14 minutes, constituting 4 minutes of slight head movement, 4 minutes of random head movements, and 2 minutes for each of rolling, yawing, and pitching. During the collection period, the participant was first asked to control their head in a steady state for 4 minutes while doing all four activities, leading to slight head movement interference on captured data. After that, the participant performed all activities casually for another 4 minutes. Head movements are unavoidable in daily life, so random interferences are added to IMU data collected. Rolling (tilting), yawing (waving), and pitching (nodding) are chosen as specific head movements, as depicted in Figure~\ref{fig:earables}. We want to evaluate the impact of each kind of head movement on the performance of cross-domain HAR tasks separately.

The collected dataset is segmented into non-overlapping windows, each encompassing sequences of 4 seconds. Each sequence corresponds to a single activity label. In consideration of the variability introduced by device orientation, we compute the magnitudes of acceleration and gyroscope data to serve as a more stable input feature set for the deep learning model. Subsequently, the processed 840 data samples are randomly partitioned into distinct subsets for model training (10\%), validation (10\%), and testing (80\%). As a result, the training set contains only 84 samples while the testing set contains 672 samples.

\subsubsection{Public Datasets}
The source domain refers to smartphone-based publicly available IMU datasets. Thus, four datasets are chosen as they are widely utilized in previous works~\cite{saeed2019multi, xu2021limu}. The four publicly available datasets comprise IMU signals gathered from individuals engaging in various activities while wearing smartphones. These datasets vary not only in the model of the smartphones used but also in the range of activities performed by the participants. Additionally, some of these datasets are further diversified by sensor locations. By incorporating these datasets into the experiment, the adaptability and effectiveness of the algorithm across diverse domains can be facilitated.

\paragraph{MotionSense} The MotionSense dataset~\cite{malekzadeh2019mobile} was collected by accelerometer and gyroscope sensors of Apple iPhone6s in 50 Hz. There are six different activities designed for the dataset \textit{(walking, walking upstairs, walking downstairs, standing, jogging and sitting)}, and the smartphones were kept in the participants' pockets of trousers.

\paragraph{HHAR} The Heterogeneity Human Activity Recognition (HHAR) dataset~\cite{stisen2015smart} contains IMU signals captured between 50-200 Hz for six different activities \textit{(walking, walking upstairs, walking downstairs, standing, biking and sitting)}. There are six models of smartphones in the dataset (three models of Samsung Galaxy and one model of LG), carried by the users around their waist.

\paragraph{UCI HAR} The UCI HAR dataset~\cite{reyes2016transition} was obtained from a waist-mounted Samsung Galaxy S2 smartphone. 6-axis IMU signals were collected under six different activities \textit{(walking, walking upstairs, walking downstairs, standing, lying down and sitting)} at 50 Hz.

\paragraph{Shoaib} The Shoaib dataset~\cite{shoaib2014fusion} consists of the motion data of seven daily activities \textit{(walking, walking upstairs, walking downstairs, standing, jogging, biking and sitting)}. During data collection, five Samsung Galaxy SII models were placed in five body positions. The sampling rate is 50 Hz.

We randomly extract samples corresponding to four activities of interest: walking, walking upstairs, standing, and jogging. A balanced dataset comprising 10,000 instances is constructed, ensuring an equal representation of 2,500 samples for each activity to prevent class imbalance during the training process. In line with the sampling frequency of the IMU signals from the Airpods, the data is down-sampled to 25 Hz. The pre-processing steps are consistent with those applied to the Airpods IMU signals, involving the segmentation of the data into non-overlapping sequences of 100 samples each. The prepared dataset is then divided, allocating 80\% for training, 10\% for validation, and the remaining 10\% for testing. Finally, the training set comprises 8,000 samples, which is 100 times larger than the training samples collected from earables.

\subsection{Overall Performance}
Table~\ref{tab:model_performance_gr_filter} summarizes the system's overall performance on the earable data test set, including both accuracy and F1-score. Figure~\ref{fig:confusion_matrix}(a) further visualizes the confusion matrix. The proposed framework demonstrates an exceptional ability to classify all four activities, regardless of the original sensor domain. The model achieves an average accuracy of 88.8\% and an F1-score of 0.89. While excelling in classifying jogging activities with a 99\% accuracy, the model exhibits lower performance for upstairs activities, sometimes misclassified as less-strong activities like walking.

\subsection{Effectiveness of domain adaptation}
To evaluate the effectiveness of the proposed domain adaptation component, keeping all the settings the same as \sysname, we discard the domain classifier and only use the 8,000 data samples from public datasets to train a classifier.
The trained model is then applied to test the AirPods samples.
As shown in Figure~\ref{fig:confusion_matrix}(b), the model's performance shows a significant decline, achieving an accuracy of only 46\%. This decline in performance clearly indicates that the model trained on public IMU datasets cannot be directly applied to earable IMU data due to a substantial domain gap across different sensor positions.
In Figure~\ref{fig:confusion_matrix}(b), we present the confusion matrix derived from the model's predictions. It can be observed that all jogging samples were wrongly classified as less intense activities such as walking or going upstairs.

In comparison, after adding the Gradient Reversal Layer in the back-propagation path of the domain classifier, the model exhibits strong potential in recognizing various activities, according to the confusion matrix in Figure~\ref{fig:confusion_matrix}(a). The model's performance on standing and jogging activities underlines its capability to discern static and highly dynamic activities with high precision. As mentioned, the amplitudes and patterns of IMU sequences while jogging show significant differences between earbuds and smartphones, which is addressed by domain adaptation. This significant improvement in performance highlights the effectiveness of bridging the domain gap and transferring knowledge even with limited earable domain data.

\subsection{Effectiveness of head movement mitigation}
To investigate the impact of pre-processing on system performance, we compare the results before and after applying low-pass filtering. The accuracy and F1-score without head movement mitigation are presented in Table~\ref{tab:model_performance_gr}, while Table~\ref{tab:model_performance_gr_filter} displays the results with head movement mitigation. The test set is further divided into 5 groups based on the magnitude of head movements: slight head movement, random head movement, roll, yaw, and pitch.

From Table~\ref{tab:model_performance_gr}, it is evident that the model's accuracy decreases when specific types of head movements are present, particularly under the investigation of rolling, yawing, or pitching. When participants make an effort to keep their heads steady during data collection (as seen in the slight head movement column), the accuracy reaches up to 93\%. However, when larger magnitude head movements are present, the accuracy degrades to approximately 85\%.

As expected, the filter effectively eliminates high-frequency components of the Airpods IMU data while preserving frequency components below 5 Hz, which are relevant to body activities. As shown in Table~\ref{tab:model_performance_gr_filter}, the model's performance in classifying all four activities under various head movement scenarios improves, with an overall accuracy increment of around 4\%. Notably, after applying the 5 Hz low-pass filter, there is no degradation in accuracy, regardless of the magnitudes or kinds of head movement interference. 
It is crucial to note that the utilization of more advanced techniques or components could potentially isolate and mitigate even low-frequency head movements from the captured motion signals. However, those approaches would inevitably introduce a higher computational overhead and might necessitate a larger amount of data for training, which is not desired here.

\section{Related Work}
Until now, deep learning models deployed in wearable devices facilitate many ubiquitous applications. Various sensors like IMUs, microphones, and wireless modules are widely embedded in mobile devices, and there are existing works focusing on human activity recognition or human interaction detection based on sensor signals captured by mobile devices, especially smartphones~\cite{jiang2015human, yao2017deepsense, liu2020giobalfusion, ni2023uncovering, chen2022swipepass, ni2023recovering, ni2023xporter}. In industry or academia, ear-worn devices have been a promising wearable technology for areas like speech enhancement~\cite{duan2024earse, he2023towards}, eating episodes detection~\cite{bedri2017earbit}, respiration rate measurement~\cite{roddiger2019towards}. Placed on the ear, earbuds equipped with the accelerometer and gyroscope enable the motion tracking of the body while they are less susceptible to motion disturbance as the upper body acts as a natural filter. They are widely utilized in the areas like human body movement classification or detection~\cite{kawsar2018earables, burgos2020ear, min2018exploring, hossain2019human, laporte2021detecting}. Furthermore, compared to traditional IMU sensors placed in other positions, earables can sense subtle head or facial movements~\cite{zhu2023char, li2022eario, gashi2021hierarchical}.

Domain adaptation, a specialized branch of transfer learning, aims to train a model using the data of the source domain and achieve high accuracy on a target dataset, which differs significantly from the source. Domain adaptation has been studied in the realm of sensing systems. It has been proven to be effective in addressing the domain shift issue in wireless sensing~\cite{zheng2019zero, jiang2018towards, gong2019metasense, ni2023exploiting, ni2021simple, sun2023flora, sun2024flora+, han2023mmsign, ni2023eavesdropping, sun2024rfegg} and wearable sensing~\cite{natarajan2016domain, akbari2019transferring, duan2023emgsense}. Few attempts have been made to utilize domain adaptation in earable sensing. Nguyen et al.~\cite{nguyen2021scalable} addressed the lack of publicly available datasets in Respiratory Symptom Detection using earables. A domain adaptation layer was implemented in~\cite{li2022adapsqa} to guide the model to learn common feature representations by aligning the distribution of both the source and target domains. Zhang et al.~\cite{zhang2023earphone} sought to use domain adaptation to minimize the training overhead while maximizing the model's adaptability to individual user characteristics. 

\section{Conclusion}
In this work, we introduce \sysname, a novel system for earable-based human activity recognition (HAR) that effectively addresses the challenges of domain adaptation and mitigates the impact of head movements. Our system leverages the rich datasets available from smartphone-based IMU sensors and adapts them for use with COTS earable devices, which are becoming increasingly prevalent. Through extensive experiments, we have demonstrated that the system can improve the classification accuracy to 88.8\% in activity recognition tasks, even with limited earable training data. The system's performance highlights the potential of domain adaptation in the realm of earable sensing and sets the stage for future work in this area.

\section*{Acknowledgment}
The work was supported by the Research Grants Council of the Hong Kong Special Administrative Region, China (Project No. CityU 21201420 and CityU 11201422), the Innovation and Technology Commission of Hong Kong (Project No. PRP/037/23FX and MHP/072/23), NSF of Shandong Province (Project No. ZR2021LZH010).

\bibliographystyle{IEEEtran}
\bibliography{citations}

\end{document}